\documentclass[12pt]{article}
\usepackage{a4,graphicx}
\begin{document}
\begin{titlepage}
\centering {\Large\bf P--violating effects in low--energy Compton
scattering}\\
 \vspace{1.5cm}

{\rm E.~Barto\v{s}}\\ {\em Dept. of Theor. Physics, Comenius University,
Bratislava, Slovak Republic}\\ \vspace{1cm}

{\rm V. Bytev, E. Kuraev}\\ {\em JINR, Dubna, Russia}\\ \vspace{2cm}

\begin{abstract}
 Parity of effects induced by one loop corrections of Standard model of
 Compton scattering are considered. We note the main effects arise from W
 contributions. Keeping in the mind scheme independent gauge invariance
 amplitude we calculated the one spin asymmetries for the case when initial
 electron is polarized and the case of circular polarized photon. They are
 of order $10^{-13}$.
\end{abstract}
\end{titlepage}

The effect of Compton scattering in SM was considered \cite{DD,AD} for the
range of the very high energies of LEP energy range. It was shown that
radiative corrections to cross--section are essential and can results about
$10$\% of contribution. In this paper was mentioned that the effect of RC is
negligible for the case of the low energies. We too considered the last case
for which analytical analysis are presented in the literature. The effect
absent in the frames of QED. Nevertheless of smallness of mentioned process,
one can use the powerful laser beam to experimentally test the process. The
main effect comes from the W axial--vector current. The reason why we don't
use Z boson is clear. In the case of W boson the ratio of vector and axial
coupling constants is 1. On the other hand in the case of Z boson is order
of $10^{-2}$.

Due to renormalizability of SM there is no dependence of the choice of the
propagator of gauge bosons and ghosts and besides of gauge invariance and
Bose symmetry present in amplitude. Besides there is no dependence of
renormalization scheme of the theory. Below we use the on mass--shell
regularization scheme for fermion self--energy and vertex function. We use
Feynman gauge W propagator. In both cases of polarized photon and electron
the interference of vector and axial--vector couplings of W boson with
fermion is essential. We start from consideration of self energy and vertex
function of fermion.

W contribution to the self-energy of fermion has the form:
\begin{equation}
\Sigma^u(p)=\frac{-ig^2}{2^6\pi^2}(1+\gamma_5)\int{\frac{d^4k}{i{\pi}^2}\gamma_\lambda
 (\hat{p}-\hat{k})\gamma_\lambda\frac{1}{(k^2-M^2){(p-k)}^2}}\;,
\end{equation}
and after the standard four dimension integration by loop momenta we
obtained for the unrenormalized self--energy operator:
\begin{equation}
\Sigma^u(p)=\frac{-ig^2}{2^5\pi^2}(1+\gamma_5)\hat{p}\left[{1\over2}L+{1\over6}
 \frac{p^2}{M^2}\right]\;.
\end{equation}
Where $L=\ln{\Lambda^2\over{M^2}}$, $\Lambda$ is the ultraviolet cut--off
and $M$ is W boson mass. Renormalization implies two subtractions: left and
right (here we have some deviation from QED scheme) and results

\begin{equation}
\Sigma^r(p)=\frac{ig^2}{6\times2^5{\pi}^2M^2}(\hat{p}-m)\hat{p}\gamma_5(\hat{p}-m)\;.
\end{equation}

Here and below we pay attention for contributions that containing $\gamma_5$.
Consider next the vertex function. It has a form
\begin{equation}
V^{\mu}(p,p+k_1)=\frac{-ieg^2}{2^6\pi^2}\int{\frac{d^4k}{i{\pi}^2}
   \frac{V_{\mu\lambda\sigma}\gamma_\sigma(p_1-k)\gamma_\lambda(1-\gamma_5)}{(012)}}\;,
\end{equation}
where $V_{\mu\lambda\sigma}$ is defined as
\begin{eqnarray}
V_{\mu\lambda\sigma}=g_{\mu\lambda}{(k-k_1)}_\sigma+g_{\lambda\sigma}{(-2k-k_1)}_\mu+
g_{\sigma\mu}{(2k_1+k)}_\lambda\;,  \\ (0)= k^2-M^2  \;,\quad
(1)={(p_1-k)}^2 \;,\quad (2)={(k+k_1)}^2-M^2 \nonumber
\end{eqnarray}

Performing the loop momentum integration and imposing the condition
${V_\mu(p,p+k_1)|}_{k_1=0}=0$ we obtain
\begin{equation}
V_\mu^r(p_1,p_1+k_1)=\frac{ieg^2}{2^6\pi^2M^2} \left[-{3\over2}\hat{k}_1\gamma_\mu m-
{17\over12}\chi_1\gamma_\mu+{7\over6}p_{1\mu}\hat{k}_1\right] , \chi_{1,2}=2p_1k_{1,2}
\end{equation}
The remaining part of diagram contribution is ultraviolet finite and may be
calculated by standard way. As a result the radiative correction to the
matrix element may be written in the form:
\begin{equation}
M_1=i{N\over2}\bar{u}(p_2)\gamma_5 O^{(1)}_{\mu\nu}u(p_1)e_{1\mu}e_{2\nu}^*,
N=\frac{e^4}{2^5\pi^2M^2\sin^2\theta_W}\;, sin^2\theta_W=0.23,
\end{equation}
where
\begin {equation}
O_{\mu\nu}^{(1)}e_{1\mu}^*=V_{1\nu}+V_{2\nu}+V_{3\nu}+V_{4\nu}+\Sigma_{\nu}+
D_{\nu}+G_{\nu}+B_{\nu},
\end{equation}
where $V_i$ represent the vertex contributions, $\Sigma$ is the self--energy
of fermion, $D$ correspond to four boson vertex diagram contribution, $G$
takes the account the ghost contribution and finally $B$ correspond to the
box contribution.

Contribution of vertex type Feynman diagram has a form (we put here only the
contributions providing the parity odd effect in the cross--section):
\begin{eqnarray}
V_{1\nu}=\frac{2p_{2\nu}+\gamma_\nu\hat{k_2}}{\chi_1}\left[-{3\over2}\hat{k}_1\hat{e}_1 m
- {17\over12}\chi_1\hat{e_1}+{7\over6}(p_1e_1)\hat{k}_1\right],\\
V_{2\nu}=\frac{2p_2e_1-\hat{e_1}\hat{k}_1}{-\chi_2}\left[{3\over2}\hat{k}_2\gamma_\nu m+
{17\over12}\chi_2\gamma_\nu-{7\over6}p_{1\nu}\hat{k}_2\right],\\
V_{3\nu}=\left[{3\over2}\gamma_\nu\hat{k}_2m-{17\over12}\gamma_\nu\chi_1+
 {7\over6}\hat{k_2}p_{2\nu}\right]\frac{(2p_1e_1)+\hat{k_1}\hat{e_1}}{\chi_1},\\
V_{4\nu}=\left[-{3\over2}\hat{e_1}\hat{k}_1m+{17\over12}\chi_2\hat{e_1}-
 {7\over6}\hat{k}_1(p_2e_1)\right]\frac{2p_{1\nu}-\hat{k}_2\gamma_\nu}{-\chi_2}.
\end{eqnarray}

Fermion self energy diagram contribution is
\begin{equation}
\Sigma_{\nu}={1\over3}\left[\gamma_\nu(\hat{p}_1+\hat{k}_1)\hat{e_1}+
   \hat{e_1}(\hat{p}_2-\hat{k}_1)\gamma_\nu\right].
\end{equation}

The $\gamma\gamma WW$ vertex containing Feynman diagram contribution is
\begin{equation}
D_\nu=e_{1\nu}\left(m+{1\over2}\hat{q}\right)+\gamma_\nu\left({(p_1e_1)}+
{1\over2}(q e_1)\right)+\hat{e_1}\left(p_{1\nu}+{1\over2}q_\nu\right),
\end{equation}
where $ q=\hat{k_1}-\hat{k_2} $.

Only two Feynman diagrams containing the ghost particle contribution are
relevant (we may neglect interaction ghost with fermion, which contribution
is suppressed by additional factor $m\over M$, m is electron mass):
\begin{equation}
G_\nu={1\over2}\hat{e_1}p_{1\nu}+{1\over2}\gamma_\nu(p_1e_1)-
{1\over2}e_{1\nu}m+\gamma_\nu\left(2\hat{k}_1-\hat{k}_2\right)\hat{e_1}+
\hat{e_1}\left(\hat{k}_1-2\hat{k}_2\right)\gamma_\nu.
\end{equation}

At least the contribution of Feynman diagrams containing three boson
vertices has a form:
\begin{eqnarray}
B_\nu=\frac{1}{24}\left[e_{1\nu}\left(-32m-28\hat{q}\right)+
 \gamma_\nu\left(35\hat{k}_2+23\hat{k}_1\right)\hat{e_1}+\right.\nonumber\\
 +\left.\hat{e_1}\left(-35\hat{k}_1-23\hat{k}_2\right)\gamma_\nu+
 28\hat{e_1}p_{1\nu}+26\hat{e_1}k_{1\nu}+
 28\gamma_\nu(p_1e_1)-26\gamma_\nu(k_2e_1)\right]
\end{eqnarray}

Parity odd contributions to cross--section arise from interference of the
Born matrix element
$M^{(0)}=ie^2\bar{u}(p_2)O^{(0)}_{\mu\nu}u(p_1)e_{1\mu}e_{2\nu}^*$ with the
one loop corrected matrix element given above. This summed on polarization
states of electrons and photons Born matrix element squared is known
proportional to:
\begin{eqnarray}
\sum{|M^{(0)}|}^2&\sim&{1\over4}Sp(\hat{p}_2+m)O^{(0)}(\hat{p}_1+m)\tilde{O}^{(0)}=S=\nonumber\\
&=&2\left(\frac{\chi_1}{\chi_2}+\frac{\chi_2}{\chi_1}\right)+8m^2\left(\frac{1}{\chi_1}
  -\frac{1}{\chi_2}\right)+8m^4{\left(\frac{1}{\chi_1}-\frac{1}{\chi_2}\right)}^2.
\end{eqnarray}
In the case of polarized electron we use it's density matrix:
$u(p,a)\bar{u}(p,a)=(\hat{p}+m)(1-\gamma_5\hat{a})$. Corresponding
interference matrix has a form:
\begin{eqnarray}
S_1&=&{1\over4}Sp(\hat{p}_2+m)\gamma_5O^{(1)}(\hat{p}_1+m)(-\gamma_5\hat{a})\tilde{O}=\nonumber\\
&=&(k_1a)m\left[-6\frac{\chi_2}{\chi_1}-\frac{7m^2}{3\chi_1}+
  7\frac{m^2}{\chi_2}-1\right]+ \\
&+&(k_2a)m\left[6\frac{\chi_1}{\chi_2}+7\frac{m^2}{\chi_1}-
\frac{7m^2}{3\chi_2}+1\right]. \nonumber
\end{eqnarray}
In the laboratory frame $\chi_1=2m\omega_1$,  $\chi_2=2m\omega_2$, and the
energy of scattering photon is
$\omega_2=\omega_1/\left[1+(\omega_1/m)(1-c)\right]$, $c=\cos\theta$ and
$\theta$ is the angle between the initial and scattered photon 3--momenta.
Scalar products entering $S_1$ are: $(k_1a)=-\omega_1|a|\cos\theta_0$,
$(k_2a)=-\omega_2|a|(\cos\theta\cos\theta_0-\sin\theta\sin\theta_0\cos\phi)$,
$\phi$ is the azimuthal angle between the planes containing initial photon
and electron spin and the plane containing initial and scattered photons
momenta.

For the case unpolarized electron and circularly polarized initial photon
(it's spin density matrix has a form
$e_{1,i}e_{1,j}^*={1\over2}(1+\xi_2\sigma_2)_{ij},\; i,j=x,y$) the
corresponding interference has a form
\begin{eqnarray}
S_2&=&\frac{1}{8 m^2}\chi_1{\chi_2}^2(1-c^2)\xi_2(A_2-A_1)-{1\over2}\xi_2\chi_1A_3-
 \nonumber\\
  &-&{1\over2}\xi_2\chi_2A_4+\frac{1}{4m^2}\xi_2(1-c)\chi_1\chi_2A_5
\end{eqnarray}
with
\begin{eqnarray}
A_1&=&-\frac{5m^2}{2\chi_1\chi_2}-\frac{31}{12\chi_1}+\frac{23}{4\chi_2}-
\frac{4m^2}{\chi_2^2}, \\ \nonumber
A_2&=&-\frac{m^2}{6\chi_1\chi_2}+\frac{119}{12\chi_1}+\frac{21}{4\chi_2}+\frac{4m^2}{3\chi_2^2},\\
\nonumber
A_3&=&\frac{29}{12}-\frac{4\chi_1}{3\chi_2}+\frac{5\chi_1m^2}{4\chi_2^2}+\frac{25\chi_2}{8\chi_1}
     +\frac{25m^2}{6\chi_1}-\frac{11\chi_2m^2}{6\chi_1^2}-\frac{29m^2}{12\chi_2},
     \\ \nonumber
A_4&=&\frac{5}{12}+\frac{\chi_1}{3\chi_2}-\frac{7\chi_1m^2}{12\chi_2^2}-\frac{31\chi_2}{24\chi_1}
     -\frac{7m^2}{12\chi_1}, \\ \nonumber
A_5&=&-\frac{1}{6}-\frac{4\chi_1}{3\chi_2}+\frac{5\chi_1m^2}{4\chi_2^2}-\frac{31\chi_2}{24\chi_1}
     +\frac{7m^4}{6\chi_1\chi_2}-\frac{5m^2}{12\chi_1}+\frac{14m^2}{\chi_2}+
     \frac{7m^4}{6\chi_2^2}.
\end{eqnarray}
The corresponding asymmetries are:
\begin{eqnarray}
{\cal{A}}_1&=&N\frac{S_1}{Sm^2}=\frac{d\sigma(\vec{a},\theta,y)-d\sigma(-\vec{a},\theta,y)}
{d\sigma(\vec{a},\theta,y)+d\sigma(-\vec{a},\theta,y)}, \\ \nonumber
{\cal{A}}_2&=&N\frac{S_2}{Sm^2}=\frac{d\sigma_R(\theta,y)-d\sigma_L(\theta,y)}
{d\sigma_R(\theta,y)+d\sigma_L(\theta,y)}, \\
N&=&\frac{g^2m^2}{32\pi^2M^2}\approx\frac{\alpha m^2}{2\pi M^2}=4\times
10^{-13}, y=\omega_1/m. \nonumber
\end{eqnarray}

In Thomson limit $y\rightarrow 0$ we have:
\begin{eqnarray}
\frac{S_1}{Sm^2}&=&\frac{7}{6}\frac{(\vec{n}_1+\vec{n}_2)\vec{a}}{(1+c^2)},\\
\frac{S_2}{Sm^2}&=&\frac{7\xi_2(1-c)}{24(1+c^2)},
\end{eqnarray}
$\vec{n}_1$, $\vec{n}_2$ are the orts along initial and the scattered
photons momenta, $\vec{a}$ is the polarization vector of the initial
electron, $\xi_2$ is degree of circular polarization of initial photon.

We are grateful to V. G. Baryshevski for suggesting the problem and to Ya. Azimov
, M. Visotsky for discussions.


\begin{thebibliography}{99}
\bibitem{DD}
Dittmaier S.: Nucl. Phys. {\bf B423} (1994) 384.
\bibitem{AD}
Denner A., Dittmaier S.: Nucl. Phys. {\bf B407} (1993) 43.
\end{thebibliography}
\end{document}